\documentclass[aps,prb,twocolumn,superscriptaddress,showpacs]{revtex4}
\usepackage{amssymb}
\usepackage{amsmath}
\usepackage{amsfonts}
\usepackage{mathrsfs}
\usepackage{epsfig}
\usepackage{natbib}
\usepackage{amsmath}
\usepackage{graphicx}
\usepackage{color}
\usepackage{amscd}

\begin{document}

\title{Probing the quantum behaviors of a nanomechanical resonator coupled
to a double quantum dot}
\author{Zeng-Zhao Li}
\affiliation{Department of Physics, State Key Laboratory of Surface
Physics, Fudan University, Shanghai 200433, China}
\affiliation{Department of Applied Physics, Hong Kong Polytechnic
University, Hung Hom, Hong Kong, China}
\author{Shi-Hua Ouyang}
\affiliation{Department of Physics, State Key Laboratory of Surface
Physics, Fudan University, Shanghai 200433, China}
\affiliation{Department of Applied Physics, Hong Kong Polytechnic
University, Hung Hom, Hong Kong, China}
\author{Chi-Hang Lam}
\affiliation{Department of Applied Physics, Hong Kong Polytechnic University, Hung Hom,
Hong Kong, China}
\author{J. Q. You}
\altaffiliation[Electronic address:~jqyou@fudan.edu.cn]{}
\affiliation{Department of Physics, State Key Laboratory of Surface
Physics, Fudan University, Shanghai 200433, China}
\affiliation{Beijing Computational Science Research Center, Beijing
100084, China}
\date{\today}

\begin{abstract}
We propose a current correlation spectrum approach to probe the
quantum behaviors of a nanomechanical resonator (NAMR). The NAMR is
coupled to a double quantum dot (DQD), which acts as a quantum
transducer and is further coupled to a quantum-point contact (QPC).
By measuring the current correlation spectrum of the QPC, shifts in
the DQD energy levels, which depend on the phonon occupation in the
NAMR, are determined. Quantum behaviors of the NAMR could, thus, be
observed. In particular, the cooling of the NAMR into the quantum
regime could be examined. In addition, the effects of the coupling
strength between the DQD and the NAMR on these energy shifts are
studied. We also investigate the impacts on the current correlation
spectrum of the QPC due to the backaction from the charge detector
on the DQD.
\end{abstract}

\maketitle

\section{Introduction \label{sec:intro}}

The observation of quantum-mechanical behaviors in
nanoelectromechanical systems, in particular, nanomechanical
resonators (NAMRs) for testing
the basic principles of quantum mechanics \cite%
{Blencowe2004PhysRep,Schwab2005PhysToday,Cho2003Science} has become
a topic of considerable interest and activity. Besides their wide
range of potential applications,
\cite{Blencowe2004PhysRep,Schwab2005PhysToday} e.g., serving as
ultrasensitive sensors in high-precision displacement measurements,
and detection of gravitational waves, quantized NAMRs are
potentially useful for quantum-information processing. For example,
NAMRs may serve as a unique intermediary for transferring quantum
information between microwave and optical domains because they can
be coupled to electromagnetic waves of any frequency.
\cite{Regal2011JPhysConf}


At very low temperatures (in the milli-Kelvin range), NAMRs of
high-vibration frequencies (gigahertz range) have recently been
experimentally verified to
approach the quantum limit. \cite%
{Gaidarzhy2005PRL,Badzey-Mohanty2005Nature,Hensinger2005PRA,LaHaye2004Science,Huang2003Nature,Knobel-Cleland2003Nature,Cleland2002APL}
However, low-frequency ($\lesssim100$~MHz) mechanical oscillators
have the distinct advantages of high-quality factors, long phonon
lifetimes, and large motional state displacements, which are
important for future testing of quantum theory
\cite{Marshall-Bouwmeester2003PRL} and other applications. A
formidable challenge (see, e.g., Refs.~%
\onlinecite{Gaidarzhy2005PRL,Badzey-Mohanty2005Nature,Hensinger2005PRA,LaHaye2004Science,Huang2003Nature,Knobel-Cleland2003Nature}%
) in this field is to detect the quantum quivering (zero-point
motion) of an NAMR so as to quantitatively verify whether it has
been cooled into the quantum-mechanical regime or not. To directly
detect the extremely small displacements of an NAMR vibrating at
gigahertz frequencies by using available
displacement-detection techniques is very difficult. \cite%
{LaHaye2004Science,Huang2003Nature,Knobel-Cleland2003Nature,Cleland2002APL}
The usual position-measurement method is also severely limited by the
``zero-point displacement" fluctuations in the quantum regime, \cite%
{Bocko1996RMP} although near-Heisenberg-limited measurements have
been performed in recent experiments.\cite{Teufe2011Nature}

In this paper, we propose a current spectroscopic approach to study
the behaviors of an NAMR. It is based on the detection of the
current correlation spectrum in a charge detector, e.g., a
quantum-point contact (QPC), which is indirectly coupled to an NAMR
via a double quantum dot (DQD)
acting as a quantum-electro-mechanical transducer. \cite%
{Geller-Cleland2005PRA-Sun2006PRA} Based on this proposal, we show
that one can observe the quantum behaviors of the NAMR and can
further verify whether it has been cooled into the quantum regime.
In contrast to a previous approach based on the superconducting
qubit coupled to a cavity, which involves Rabi splitting,
\cite{Wei2006PRL} our proposed setup is expected to provide better
tunability via the gate voltages. Moreover, we also study the
effects of the backaction from the charge detector on the DQD.

We consider a coupled NAMR-DQD system in the strong \textit{dispersive regime%
} where the coupling strength $g$ is much smaller than the frequency
detuning $\delta $ between the DQD and the NAMR. In this regime,
energy quanta in the NAMR are only virtually exchanged between the
DQD and the NAMR. Thus, the coupling of the DQD to the NAMR does not
change the occupation state of the electron in the DQD, but only
results in phonon-number-dependent frequency shifts in the DQD
energy levels. These shifts are analogous to Stark shifts and can be
further detected by measuring the current correlation spectrum of
the QPC.

This paper is organized as follows. In Sec.~\ref{sec:model}, the coupled
system is explained. The effective dispersive Hamiltonian is derived in Sec.~%
\ref{sec:effective Hamiltonian}. The quantum dynamics of the coupled
NAMR-DQD system in the presence of the QPC are derived in Sec.~\ref%
{sec:master equation}. In Sec.~\ref{sec:spectrum}, results related
to the observation of quantum behaviors, the verification of the
ground-state cooling of an NAMR, as well as the backaction from the
QPC on the DQD are analyzed. Conclusions are given in
Sec.~\ref{conclusions}.

\section{The coupled NAMR-DQD-QPC system \label{sec:model}}

\begin{figure}[b]
\centerline{\includegraphics[width=3.2in,bbllx=89,bblly=82,bburx=557,bbury=503]
{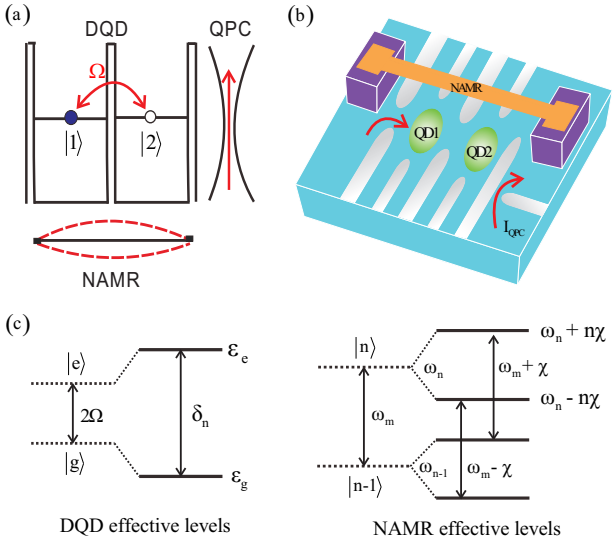}} \caption{(Color online)~(a)~Schematic of an NAMR
capacitively coupled to a DQD, which is under measurement by a
nearby QPC. The energy detuning between the two dot states in the
DQD is zero, and the interdot coupling strength between them is
$\Omega$. (b)~A stereographical diagram of the device where an
electron is injected from the left reservoir to the DQD by changing
the gate voltages, and the displacement of the metallic NAMR from
its equilibrium position modulates the capacitance between the NAMR
and the DQD. (c)~Effective energy levels for the DQD (horizontal
solid lines in the left panel) in the dispersive DQD-NAMR coupling
regime:
$\protect\varepsilon_e=\Omega+\left(n+1/2\right)\protect\chi$ and
$\protect\varepsilon_g=-\Omega-\left(n+1/2\right)\protect\chi$ with
energy detuning
$\protect\delta_n=2\Omega+\left(2n+1\right)\protect\chi$; effective
energy levels for the NAMR (horizontal solid lines in the right
panel) in the dispersive DQD-NAMR coupling regime:
$\protect\omega_n+n\protect\chi$ and $\protect\omega_n-n
\protect\chi$ with $\protect\omega_n=n\protect\omega_m$. The
effective phonon level differences are $\protect\omega_m +\protect\chi$ and $\protect\omega%
_m -\protect\chi$.}
\label{NAMR-DQD-QPC model}
\end{figure}

The device layout of an NAMR capacitively coupled to a lateral DQD,
which is further measured by a QPC, is presented in
Fig.~\ref{NAMR-DQD-QPC model}(a).
Here, we consider a Coulomb-blockade regime with strong intradot and
interdot Coulomb interactions so that only one electron is allowed
in the DQD. The states of the DQD are denoted by occupation states
$|1\rangle $ and $|2\rangle $, representing one electron in the left
and the right dots, respectively. The stereographical diagram of
this device is shown in Fig.~\ref{NAMR-DQD-QPC model}(b). The
lateral DQD is formed by properly tuning the voltages applied to the
gates. Also, an electron can be injected from the left reservoir to
the DQD by changing the gate voltages. A metallic NAMR is fabricated
above the DQD, and the displacement of the NAMR from its equilibrium
position modulates the mutual capacitance between the NAMR and the
DQD. \cite{You2009PRB} The current $I_{\mathrm{QPC}}$ through the
QPC depends on the electron occupation of the DQD.

The total Hamiltonian of the whole system is
\begin{equation}
H=H_{\mathrm{sys}}+H_{\mathrm{int}}+H_{\mathrm{\det }},
\label{eq:rotation_hamiltonian}
\end{equation}%
with an unperturbed part,
\begin{equation}
H_{\mathrm{sys}}=H_{\mathrm{NAMR}}+H_{\mathrm{DQD}}+H_{\mathrm{QPC}},
\end{equation}%
where (after setting $\hbar =1$),
\begin{equation}
H_{\mathrm{NAMR}}=\omega _{m}b^{\dag }b,  \label{eq:NAMR_hamiltonian}
\end{equation}%
\begin{equation}
H_{\mathrm{DQD}}=\frac{\Delta }{2}(a_{2}^{\dag }a_{2}-a_{1}^{\dag
}a_{1})+\Omega (a_{2}^{\dag }a_{1}+a_{1}^{\dag }a_{2}),
\label{eq:DQD_hamiltonian}
\end{equation}%
\begin{equation}
H_{\mathrm{QPC}}=\sum_{k}\omega _{Sk}c_{Sk}^{\dag }c_{Sk}+\sum_{q}\omega
_{Dq}c_{Dq}^{\dag }c_{Dq}.  \label{eq:QPC_hamiltonian}
\end{equation}%
The interaction parts are
\begin{equation}
H_{\mathrm{int}}=-(g_{2}a_{2}^{\dag }a_{2}+g_{1}a_{1}^{\dag }a_{1})\left(
b^{\dag }+b\right) ,
\end{equation}%
\begin{equation}
H_{\mathrm{\det }}=\sum_{kq}\left( T_{0}-\zeta _{2}a_{2}^{\dag }a_{2}-\zeta
_{1}a_{1}^{\dag }a_{1}\right) \left( c_{Sk}^{\dag }c_{Dq}+c_{Dq}^{\dag
}c_{Sk}\right) .  \label{eq:QPC-DQD_hamiltonian}
\end{equation}%
Here $H_{\mathrm{NAMR}}$, $H_{\mathrm{DQD}}$, and
$H_{\mathrm{QPC}}$, respectively, are the free Hamiltonians of the
NAMR, the DQD, and the QPC without the tunneling terms. The phonon
operators $b^{\dagger }$ and $b$, respectively, create and
annihilate an excitation of frequency $\omega _{m}$ in the NAMR. In
Eq.~(\ref{eq:DQD_hamiltonian}), $\Delta $ is the energy detuning
between the two dots, and $\Omega $ is the interdot coupling. Below,
we consider, for simplicity, the degenerate-state case with $\Delta
=0$ [see Fig.~\ref{NAMR-DQD-QPC model}(a)]. $c_{Sk}$ ($c_{Dq}$) is
the annihilation operator for electrons in the source (drain)
reservoir of the QPC with momentum $k$ ($q$). Here, we define
pseudospin operators $\sigma _{z}\equiv a_{2}^{\dag
}a_{2}-a_{1}^{\dag }a_{1}$ and $\sigma _{x}\equiv a_{2}^{\dag
}a_{1}+a_{1}^{\dag }a_{2}$ with $a_{1}$ ($a_{2}$) being the
annihilation operator for an electron staying at the left (right) dot. $H_{%
\mathrm{int}}$ is the electromechanical coupling between the NAMR
and dots 1 and 2 with coupling strengthes $g_{1}$ and $g_{2}.$ The
relative coupling strengths $g\equiv (g_{2}-g_{1})/2$ is about
$0.1\omega _{m}\sim 0.5\omega
_{m}$ for typical electromechanical couplings (see, e.g., Ref.~%
\onlinecite{LambertNori2008PRB}). $H_{\mathrm{det}}$ describes
tunnelings in the QPC, which depends on the electron occupation of
the DQD, owing to the electrostatic coupling between the DQD and the
QPC. We define $T\equiv T_{0}-(\zeta _{2}+\zeta _{1})/2$ and $\zeta
\equiv (\zeta _{2}-\zeta _{1})/2$ so that the transition amplitude
of the QPC when an extra electron stays at the left and right dots
equals $T+\zeta $ or $T-\zeta $, respectively.

\section{Effective dispersive Hamiltonian \label{sec:effective Hamiltonian}}

In the eigenstate basis, the DQD Hamiltonian can be written as%
\begin{equation}
H_{\mathrm{DQD}}=\Omega \varrho _{z},
\end{equation}%
where $\varrho _{z}=a_{e}^{\dag }a_{e}-a_{g}^{\dag }a_{g}$ with the
two eigenstates of the DQD given by $\left\vert g\right\rangle
=\left( \left\vert 1\right\rangle -\left\vert 2\right\rangle \right)
/\sqrt{2}$ and $\left\vert e\right\rangle =\left( \left\vert
1\right\rangle +\left\vert 2\right\rangle \right) /\sqrt{2}$ and the
energy splitting between these two eigenstates is $2\Omega$. Then,
the total Hamiltonian becomes
\begin{eqnarray}
H &=&\omega _{m}b^{\dag }b\,+\Omega \varrho _{z}+g\varrho
_{x}\left( b^{\dag }+b\right)  \notag \\
&&+H_{\mathrm{QPC}}+\sum_{kq}\left[ T+\zeta \varrho _{x}\right]
\left( c_{Sk}^{\dag }c_{Dq}+c_{Dq}^{\dag }c_{Sk}\right)
,\label{eq:effective general_hamiltonian}
\end{eqnarray}%
where $\varrho _{x}=a_{e}^{\dag }a_{g}+a_{g}^{\dag }a_{e}.$

In the dispersive DQD-NAMR coupling regime with $|\eta|<1,$ where
$\eta =g/\delta$ and $\delta =2\Omega -\omega_{m},$ applying a
canonical transformation $UHU^{\dag }$ on the Hamiltonian
(\ref{eq:effective general_hamiltonian}), where $U=e^{s}$ with
$s=\eta \left( \varrho _{+}b-\varrho _{-}b^{\dag }\right) ,$  one
obtains an effective dispersive Hamiltonian. Under the rotating-wave
approximation, this dispersive Hamiltonian can be written, up to
second
order in $\eta ,$ as $H=H_{0}+H_{\mathrm{I}},$ with%
\begin{equation}
H_{0}=\omega _{m}b^{\dag }b\,+\Omega \varrho _{z}+\chi \left( \frac{1}{2}%
+b^{\dag }b\right) \varrho _{z}+H_{\mathrm{QPC}},
\label{eq:effective_hamiltonian}
\end{equation}%
\begin{equation}
H_{\mathrm{I}}=\sum_{kq}\left( T+\zeta \varrho _{x}\right) \left(
c_{Sk}^{\dag }c_{Dq}+c_{Dq}^{\dag }c_{Sk}\right) .
\label{eq:effective_interaction_hamiltonian}
\end{equation}%
Here, $\chi =g^{2}/\delta $. The third term in
Eq.~(\ref{eq:effective_hamiltonian}) is a dispersive interaction
that can be viewed as either a DQD-state-dependent frequency shift
in the NAMR or a phonon-number-dependent shift in the DQD transition
frequency. This interaction implies that, when the DQD state is
excited (deexcited), an energy $2\chi $ is effectively added to
(removed from) each NAMR phonon. A similar frequency shift also
appears in analogous systems in quantum optics.
\cite{Scully-Zubairy} The dispersive NAMR-DQD
energy levels, described by the first three terms in Eq.~(\ref%
{eq:effective_hamiltonian}), are the quantum version of the ac Stark
effect. When there is no interaction ($g=0$) between the NAMR and
the DQD, energy differences between adjacent levels of the NAMR or
the DQD are simply $\omega _{m}$ or $2\Omega $, respectively.
However, for $g>0,$ these eigenstates are dressed by the dispersive
interaction. The corresponding phonon level
differences become $\omega _{m}-\chi $ for the DQD state $|g\rangle $ and $%
\omega _{m}+\chi $ for state $|e\rangle ,$ whereas the DQD energy split is%
\begin{equation}
\delta _{n}\equiv 2\Omega +(2n+1)\chi
\end{equation}%
for phonon number $n$ in the NAMR. Figure~\ref{NAMR-DQD-QPC
model}(c) shows these effective energy-level differences. The
phonon-number-dependent frequency shift in the DQD as well as the
DQD-state-dependent shift in the NAMR can be detected as will be
explained below.

\section{Quantum dynamics of the coupled NAMR-DQD system \label{sec:master
equation}}

We now derive a master equation to describe the quantum dynamics of
the coupled system. In the interaction picture with the dispersive Hamiltonian $%
H_{0}$ in Eq.~(\ref{eq:effective_hamiltonian}), the interaction Hamiltonian $%
H_{\mathrm{I}}$ [Eq.~(\ref{eq:effective_interaction_hamiltonian})] can be
written as
\begin{equation}
H_{\mathrm{I}}\left( t\right) =S\left( t\right) Y\left( t\right) ,
\label{interaction_hamiltonian}
\end{equation}%
with
\begin{equation}
S\left( t\right) =\sum_{j=1}^{3}P_{j}e^{i\widehat{\omega }_{j}t},
\end{equation}%
\begin{equation}
Y\left( t\right) =\sum_{kq}\big[F_{kq}^{\dag }\left( t\right) +F_{kq}\left(
t\right) \big],
\end{equation}%
where $P_{1}=\chi \varrho _{+},$ $P_{2}=\chi \varrho _{-},$ $P_{3}=T,$ $%
\widehat{\omega }_{1}=2\Omega +2\chi \left( \frac{1}{2}+b^{\dag }b\right) ,$
$\widehat{\omega }_{2}=-\widehat{\omega }_{1},$ $\widehat{\omega }_{3}=0,$ $%
F_{kq}^{\dag }\left( t\right) =c_{Sk}^{\dag }c_{Dq+}e^{i\left(
\omega _{Sk}-\omega _{Dk}\right) },$ and $F_{kq}\left( t\right)
=c_{Dq}^{\dag }c_{Sk}e^{-i\left( \omega _{Sk}-\omega _{Dk}\right)
}.$ Applying the Born-Markov approximation and tracing over the
degrees of freedom of the QPC, quantum dynamics of the NAMR-DQD
system are governed by
\begin{eqnarray}
\dot{\rho}^{I}\left( t\right)  &=&\mathrm{Tr}_{S,D}\big\{-i\left[ H_{\mathrm{%
I}}\left( t\right) ,\rho _{\mathrm{tot}}\left( 0\right) \right]   \notag \\
&&-\int_{0}^{\infty }\left[ H_{\mathrm{I}}\left( t\right) ,\left[ H_{\mathrm{%
I}}\left( t^{\prime }\right) ,\rho _{\mathrm{tot}}\left( t\right) \right] %
\right] \big\}.  \label{equation}
\end{eqnarray}%
Here, $\rho _{\mathrm{tot}}\left( t\right) $ is the density operator
of the whole system including the QPC as well. Substituting
$H_{\mathrm{I}}$ from Eq.~(\ref{interaction_hamiltonian}) into
Eq.~(\ref{equation}) and converting the resulting equation into the
Schr\"{o}dinger picture, we obtain the master equation,
\begin{equation}
\dot{\rho}\left( t\right) =\mathcal{L}\rho \left( t\right) =-i\left[ H_{%
\mathrm{DQD}},\rho \left( t\right) \right] +\mathcal{L}_{d}\rho \left(
t\right) +\gamma _{d}\mathcal{D}\left[ \varrho _{-}\right] \rho \left(
t\right) ,  \label{ME}
\end{equation}%
with
\begin{eqnarray}
\mathcal{L}_{d}\rho \left( t\right)  &=&\big\{\sum_{i=1}^{3}\mathcal{D}\left[
P_{i}\right] \rho \left( t\right) +\sum_{i=1}^{3}\sum_{j=1\left( j\neq
i\right) }^{3}\mathcal{D}\left[ P_{i},P_{j}\right] \rho \left( t\right) %
\big\}  \notag \\
&&\times 2\pi g_{S}g_{D}\zeta ^{2}\big[\Theta \left( eV_{\mathrm{QPC}%
}-\omega _{i}\right)   \notag \\
&&+\Theta \left( -eV_{\mathrm{QPC}}-\omega _{i}\right) \big],
\end{eqnarray}%
where $\Theta \left( x\right) =\left( \left\vert x\right\vert
+x\right) /2$ and $g_{S, D}$ denotes the density of states in the source and drain reservoirs of the QPC, which has a bias voltage $%
V_{\mathrm{QPC}}.$ $\omega _{i}$ is the eigenvalue of the operator $\widehat{%
\omega }_{i}$ with the NAMR in the $\left\vert n\right\rangle $
state. Here, for simplicity, the temperatures of the reservoirs in
the DQD-QPC system (instead of the temperature $T_{m}$ of the NAMR)
are chosen to be $T=0$K because related quantum-dot
experiments are performed at extremely low temperatures (see, e.g., Ref.~%
\onlinecite{Gustavsson2007PRL}). The superoperator $\mathcal{D}$,
acting on any single or double operators, is defined as
\begin{equation}
\mathcal{D}\left[ A\right] \rho \equiv A\rho A^{\dag }-\frac{1}{2}A^{\dag
}A\rho -\frac{1}{2}\rho A^{\dag }A,
\end{equation}%
\begin{equation}
\mathcal{D}\left[ A,B\right] \rho \equiv \frac{1}{2}\big(A\rho B^{\dag
}+B\rho A^{\dag }-B^{\dag }A\rho -\rho A^{\dag }B\big).
\end{equation}%
To account for the coupling of the DQD to other degrees of freedom,
such as hyperfine interactions and electron-phonon couplings, we
have phenomenologically included an additional relaxation term [the
third term on the right-hand side of Eq.~(\ref{ME})] describing
transitions from excited state $\left\vert e\right\rangle $ to
ground state $\left\vert g\right\rangle .$ \cite{You2010PRB} In the
strong dispersive regime, as we have mentioned before, the phonon in
the NAMR neither is absorbed nor induces any transitions in the DQD
and, hence, does not change the occupation probability of the DQD.
Instead, the occupation state of the DQD is only affected by the
backaction of the QPC and the phenomenological relaxation term.

In the basis $\left\{ \left\vert e,n\right\rangle ,\left\vert
g,n\right\rangle \right\} $ of the coupled NAMR-DQD system, we obtain the
following evolution equations for the reduced density matrix elements:
\begin{equation}
\dot{\rho}_{en,en}=\gamma _{+}\rho _{gn,gn}-\left( \gamma _{-}+\gamma
_{d}\right) \rho _{en,en},  \label{ee}
\end{equation}%
\begin{equation}
\dot{\rho}_{gn,gn}=-\gamma _{+}\rho _{gn,gn}+\left( \gamma _{-}+\gamma
_{d}\right) \rho _{en,en},  \label{gg}
\end{equation}%
\begin{equation}
\dot{\rho}_{en,gn}=-i\delta _{n}\rho _{en,gn}-\gamma _{1}\left( \rho
_{en,gn}-\rho _{gn,en}\right) -\frac{1}{2}\gamma _{d}\rho _{en,gn},
\label{eg}
\end{equation}%
\begin{equation}
\dot{\rho}_{gn,en}=i\delta _{n}\rho _{gn,en}+\gamma _{1}\left( \rho
_{en,gn}-\rho _{gn,en}\right) -\frac{1}{2}\gamma _{d}\rho _{gn,en}.
\label{ge}
\end{equation}%
Assuming $eV_{\mathrm{QPC}}>\delta _{n}>0,$ the QPC-induced relaxation and
excitation rates between the ground state and the excited state of the DQD
are defined as $\gamma _{+}=\gamma _{1}\left( 1-\lambda _{n}\right) ,$ $%
\gamma _{-}=\gamma _{1}\left( 1+\lambda _{n}\right) ,$ where $\gamma
_{1}=2\pi g_{S}g_{D}\chi ^{2}eV_{\mathrm{QPC}}$ and $\lambda
_{n}=\delta _{n}/eV_{\mathrm{QPC}}.$ Since the decay rate of the
NAMR is much smaller than that of the DQD, dissipations of the NAMR
are neglected (see further discussions below). In
Eqs.~$(\ref{ee})\!\!-\!\!(\ref{ge})$, the reduced density matrix
element $\rho _{in,in}(i=g,e)$ gives the occupation probability of
state $|i,n\rangle $ of the coupled NAMR-DQD system, whereas $\rho
_{in,jn}(i\neq j)$ describes the quantum coherence between states $%
|i,n\rangle $ and $|j,n\rangle $. Equations of motion for other
elements, e.g., $\rho _{in,jn^{\prime }}(n\neq n^{\prime })$, which
are decoupled from those considered here, are not shown. Using the
normalization condition $p_{n}=\rho _{gn,gn}+\rho _{en,en}$, the
solutions to the equations above are obtained as
\begin{equation}
\rho _{en,en}\left( t\right) =\left[ \rho _{en,en}\left( 0\right) -\frac{%
\gamma _{+}}{2\gamma _{0}}p_{n}\right] e^{-2\gamma _{0}t}+\frac{\gamma _{+}}{%
2\gamma _{0}}p_{n},  \label{sol_ee}
\end{equation}%
\begin{equation}
\rho _{gn,gn}\left( t\right) =\left[ \rho _{gn,gn}\left( 0\right) -\frac{%
\gamma _{-}+\gamma _{d}}{2\gamma _{0}}p_{n}\right] e^{-2\gamma _{0}t}+\frac{%
\gamma _{-}+\gamma _{d}}{2\gamma _{0}}p_{n},  \label{sol_gg}
\end{equation}%
\begin{eqnarray}
\rho _{en,gn}\left( t\right) &=&e^{-\gamma _{0}t}\Big [\cos \left( \nu
_{n}t\right) \rho _{en,gn}\left( 0\right) +\sin \left( \nu _{n}t\right)
\notag \\
&&\times \frac{\gamma _{1}\rho _{gn,en}\left( 0\right) -i\delta _{n}\rho
_{en,gn}\left( 0\right) }{\sqrt{\delta _{n}^{2}-\gamma _{1}^{2}}}\Big ],
\label{sol_eg}
\end{eqnarray}%
\begin{eqnarray}
\rho _{gn,en}\left( t\right) &=&e^{-\gamma _{0}t}\Big [\cos \left( \nu
_{n}t\right) \rho _{gn,en}\left( 0\right) +\sin \left( \nu _{n}t\right)
\notag \\
&&\times \frac{\gamma _{1}\rho _{en,gn}\left( 0\right) +i\delta _{n}\rho
_{gn,en}\left( 0\right) }{\sqrt{\delta _{n}^{2}-\gamma _{1}^{2}}} \Big ],
\label{sol_ge}
\end{eqnarray}%
where%
\begin{equation}
\gamma _{0}=\gamma _{1}+\frac{\gamma _{d}}{2},
\end{equation}%
\begin{equation}
\nu _{n}=\sqrt{\delta _{n}^{2}-\gamma _{1}^{2}},  \label{resonant_frequency}
\end{equation}%
and $p_{n}$ is the probability that the NAMR is in state $|n\rangle
$. In the calculation, we have assumed $0<\gamma _{1}<\delta _{n}$
(see typical parameters listed in
Sec.~\ref{sec:quantum-to-classical}).


\section{Current correlation spectrum of the QPC \label{sec:spectrum}}

The dc current through the QPC is given by \cite{Gurvitz1997PRB}%
\begin{equation}
I\left( t\right) =I_{l}\rho _{11}\left( t\right) +I_{r}\rho _{22}\left(
t\right) ,
\end{equation}%
where $I_{l}=eD$ and $I_{r}=eD^{\prime }$ are the currents through
the QPC when dots $1$ and $2$, respectively, are occupied.
\cite{Gurvitz1997PRB} Here,
$D=2\pi g_{S}g_{D}\left( T-\zeta \right) ^{2}eV_{\mathrm{QPC}}$ and $%
D^{\prime }=2\pi g_{S}g_{D}\left( T+\zeta \right) ^{2}eV_{\mathrm{QPC}}$ are
the corresponding rates of electron tunneling through the QPC, which follows
from Eq.~(\ref{eq:QPC-DQD_hamiltonian}). Using
$\rho _{11}+\rho _{22}=1$, one can define the current operator as%
\begin{equation}
\widehat{I}\left( t\right) =I_{0}-I_{1}\sigma _{z},  \label{current operator}
\end{equation}%
with $I_{0}=\frac{e}{2}\,\left( D+D^{\prime }\right) $ and $I_{1}=\frac{e}{2}%
\left( D-D^{\prime }\right) $ and $\varrho _{x}=-\sigma _{z}$ in the
degenerate-state case with $\Delta=0.$ According to the
Wiener-Khintchine theorem, when the phonon in the NAMR is in state
$|n\rangle ,$ the QPC current correlation power spectrum
$S_{n}\left( \omega \right) $ is given in terms of the two-time
correlation function as \cite{Scully-Zubairy}%
\begin{eqnarray}
S_{n}\left( \omega \right) &=&2\Re \int\limits_{0}^{+\infty }d\tau
e^{i\omega \tau }\big[\big<\widehat{I}\left( t\right) \widehat{I}\left(
t+\tau \right) \big>_{n}  \notag \\
&&-\big<\widehat{I}\left( t+\tau \right)
\big>_{n}\big<\widehat{I}\left( t\right) \big>_{n}\big].
\label{n_spectrum}
\end{eqnarray}%
Substituting Eqs.~$(\ref{sol_ee})\!\!-\!\!(\ref{sol_ge})$ and (\ref{current operator}%
) into Eq.~(\ref{n_spectrum}) and using $S\left( \omega \right)
=S_{0}+\sum_{n}p_{n}S_{n}\left( \omega \right) ,$ we get%
\begin{eqnarray}
\frac{S\left( \omega \right) }{S_{0}} &=&1+\frac{2\gamma _{1}\gamma _{2}}{%
\gamma _{1}+\gamma _{2}}\sum_{n}p_{n}^{2}\Big\{\frac{\gamma _{0}}{\gamma
_{0}^{2}+\left( \nu _{n}-\omega \right) ^{2}}  \notag \\
&&\times \Big[1+\frac{\gamma _{1}}{\gamma _{0}}\left( 1+\frac{\omega }{\nu
_{n}}\right) +\frac{\gamma _{+}-\gamma _{-}-\gamma _{d}}{2\gamma _{0}}\frac{%
\delta _{n}}{\nu _{n}}\Big]  \notag \\
&&+\frac{\gamma _{0}}{\gamma _{0}^{2}+\left( \nu _{n}+\omega \right) ^{2}}%
\Big[1+\frac{\gamma _{1}}{\gamma _{0}}\left( 1-\frac{\omega }{\nu _{n}}%
\right)  \notag \\
&&+\frac{\gamma _{+}-\gamma _{-}-\gamma _{d}}{2\gamma _{0}}\frac{\delta _{n}%
}{\nu _{n}}\Big]\Big\}.  \label{QPCSpectrum}
\end{eqnarray}%
Here, $S_{0}=2eI_{0}$ is the current-noise background.
From Eq.~(\ref{QPCSpectrum}), one sees that the current correlation spectrum
of the QPC consists of peaks at resonance frequencies
$\omega =\pm \nu _{n}$. These peaks have width $\gamma _{0}$ and
heights increasing with the probability $p_{n}$. In particular, for
small backaction from the QPC, i.e., $\gamma _{1}\ll\delta _{n}$,
peaks are located at the resonance point $\omega =\delta
_{n}=2\Omega +(2n+1)\chi,$ admitting a shift $(2n+1) \chi$ inherited
from the phonon-number-dependent frequency shift in the DQD as
explained above. Thus, one can read out the phonon-number state of
the NAMR from these peak shifts in the current correlation spectrum
of the QPC.

\subsection{Verification of ground state cooling of the NAMR\label%
{sec:quantum-to-classical}}

\begin{figure}[b]
\centerline{\includegraphics[width=3.5in,bbllx=123,bblly=0,bburx=512,bbury=418]
{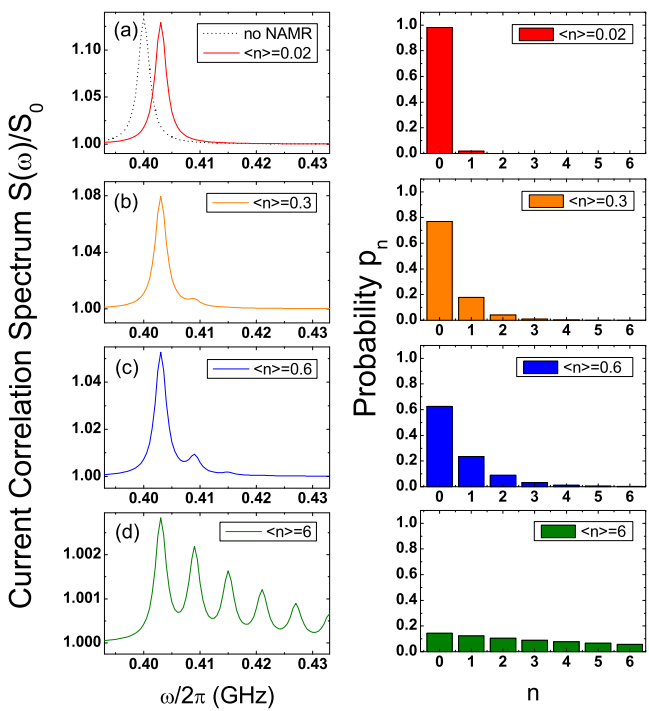}} \caption{~(Color online) Left panel:
Current correlation spectrum of the QPC when the average phonon
numbers in the NAMR are (a) $\langle n\rangle =0.02$, (b) $0.3$, (c)
$0.6$, and (d) $6$, respectively, given by the thermal
distribution, i.e., $p_{n}=\langle n\rangle ^{n}/(1+\langle n\rangle )^{n+1}$%
. Right panel: the corresponding probability of the NAMR in state $%
|n\rangle $. The coupling strength between the NAMR and the DQD is chosen as
$g=0.3~\protect\omega _{m}$. The other parameters are $\protect\omega _{m}=2%
\protect\pi \times 100$~MHz, $\Omega =2\protect\omega _{m}$, $\protect\gamma %
_{2}=0.01\protect\omega _{m}$, $\protect\gamma _{1}=0.2\protect\gamma _{2}$,
$\protect\gamma _{d}=2\protect\gamma _{2}$, and $\protect\zeta /T=0.044$.}
\label{GroundStateVerificationNAMR}
\end{figure}
The observation of quantum mechanical phenomena requires a high
frequency and a low temperature for the NAMR (see, e.g.,
Refs.~\onlinecite{Gaidarzhy2005PRL,Badzey-Mohanty2005Nature,Hensinger2005PRA})
so that $N_{\mathrm{th}}\equiv k_{\mathrm{B}}T_{m}/\hbar \omega _{m}<1$ or $%
\left\langle n\right\rangle <0.582$, where $N_{\mathrm{th}}$ is the thermal
occupation number and $k_{\mathrm{B}}$ is the Boltzmann constant. We now
assume thermal equilibrium of the NAMR with a probability \cite%
{Scully-Zubairy} $p_{n}\propto e^{-\langle n|H_{\mathrm{NAMR}}|n\rangle /k_{%
\mathrm{B}}T}$ for a state $|n\rangle $ so that $p_{n}=\left\langle
n\right\rangle ^{n}/\left( 1+\left\langle n\right\rangle \right)
^{n+1}.$ In general, a state with the average phonon number
$\left\langle n\right\rangle \ll 1$ (e.g., $\left\langle
n\right\rangle =0.01$) implying $p_{0}\approx 1$ is considered as a
quantum ground state$.$

By using typical parameters \cite%
{LambertNori2008PRB,You2010PRB,TFLi2008APL,Wiel2002RMP,vandersypen2004APL} $%
\omega _{m}=2\pi \times 100$, $\Omega =2\pi \times 200$~MHz, $%
g=0.3\omega _{m},$ $\zeta /T=0.044,$ $\gamma _{2}=0.01\omega _{m}$, $\gamma
_{1}=0.2\gamma _{2}$, and $\gamma _{d}=2\gamma _{2}$, the current
correlation spectrum of the QPC is calculated and is presented in Fig.~\ref%
{GroundStateVerificationNAMR} where only the positive frequency regime is
shown. For $\langle n\rangle =0.02\ll 0.582$ in the quantum regime [see Fig. %
\ref{GroundStateVerificationNAMR}(a)], there is only a single peak
in the spectrum corresponding to the transition frequency between
the two eigenstates of the DQD. From Eq.~(\ref{QPCSpectrum}), the
peak is located at the resonance frequency $\nu _{0}$ given in
Eq.~(\ref{resonant_frequency}). The corresponding probability
distribution function $p_{n}$ is also shown in the right panel of
the figures showing the probability of finding the NAMR in state
$\left\vert n\right\rangle $. This spectrum is nearly
indistinguishable from the pure ground state with $\langle n\rangle
=0$. At a higher temperature, other peaks begin to appear in the
current correlation spectrum
[Fig.~\ref{GroundStateVerificationNAMR}(b)]. The peak position as
given in Eqs.~(\ref{resonant_frequency}) and (\ref{QPCSpectrum})
admits an NAMR-induced shift analogous to an ac Stark shift. In the regime with, e.g., $\langle n\rangle =0.6$ [Fig. \ref%
{GroundStateVerificationNAMR}(c)] and $6$ [Fig.~\ref%
{GroundStateVerificationNAMR}(d)], multiple peaks are obtained. As
each resonance peak in the spectrum corresponds to a phonon-number
state of the NAMR, the relative area under each peak could be used,
in principle, to calculate the phonon statistics of the NAMR.

To observe multiple peaks in the correlation spectrum, the
separation between two adjacent peaks must be larger than the
intrinsic peak width, i.e., $2\chi >\gamma _{0}$. The ensemble can
then be individually resolved, which allows us to detect the phonon
number to verify the cooling efficiency of the NAMR. On the
contrary, the phonon-number state of the NAMR cannot be measured
when $2\chi <\gamma _{0}.$ Indeed, a relatively strong coupling
between an NAMR and a quantum dot has been recently demonstrated.
\cite{Huttel2009NanoLett} Thus, the regime with $2\chi >\gamma _{0}$
could be achievable. Also, the ground-state cooling of an NAMR
coupled to a DQD was proposed. \cite{You2009PRB} One can then apply
the proposed coupled NAMR-DQD system to verify the ground-state
cooling of the NAMR. The frequency shifts in the DQD energy levels
are also different for the ground and excited states of the NAMR and
can then be used to read out the phonon state of the NAMR, which can
be different from the phonon statistics in the thermal state
discussed above.

\begin{figure}[b]
\centerline{\includegraphics[width=3.2in,bbllx=95,bblly=1,bburx=423,bbury=422]
{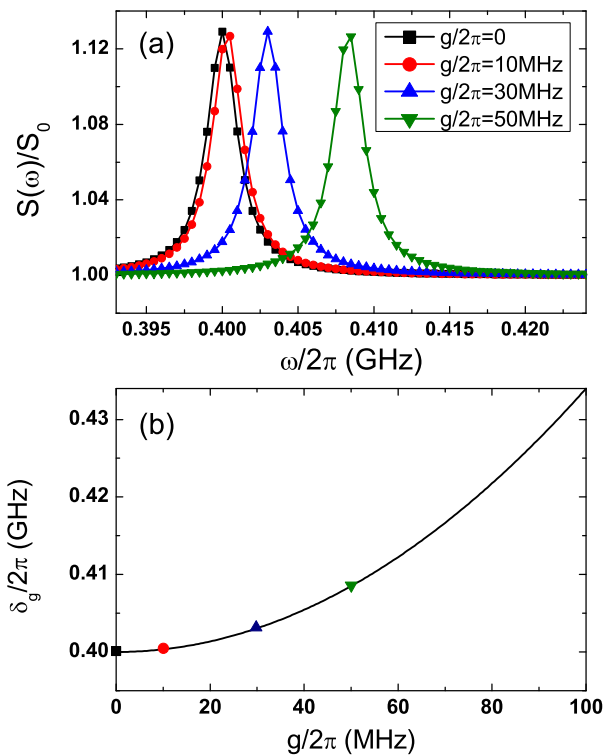}} \caption{~(Color online) (a) Current correlation
spectrum of the QPC when the NAMR is in the ground state and the
coupling strengths $g$ between the NAMR and the DQD are $0$ (black
squares), $10$~MHz (red circles), $30$~MHz (blue upper triangles),
and $50$~MHz (olive lower triangles), respectively. (b) Frequency
shift $\protect\delta_g$ as a function of the coupling strength $g$
when the NAMR is in the ground state. Other parameters are the same
as those in Fig.~\protect\ref{GroundStateVerificationNAMR}.}
\label{backactionNAMR}
\end{figure}


\begin{figure}[b]
\centerline{\includegraphics[width=3.2in,bbllx=13,bblly=14,bburx=512,bbury=392]
{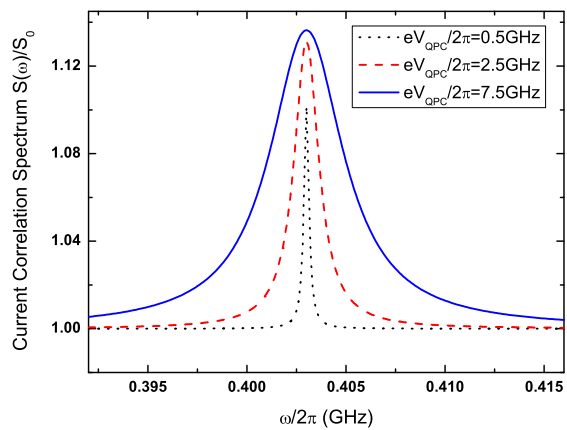}} \caption{~(Color online) Current correlation
spectrum of the QPC at various voltage biases $V_{\mathrm{QPC}}$ of
the QPC when the NAMR is approximately in the ground state, i.e.,
$\langle n \rangle=0.02$. Other parameters are the same as those in
Fig.~\protect\ref{GroundStateVerificationNAMR}.} \label{backaction}
\end{figure}

Figure~\ref{backactionNAMR}(a) shows the current correlation
spectrum of the QPC when the NAMR is in its ground state with
$\langle n\rangle =0.02$. If the NAMR and the DQD are decoupled,
i.e., $g=0$, the frequency corresponding to the peak position is
about $2\pi \times 0.4$~GHz, which is the transition frequency
$2\Omega$ between the two eigenstates of the DQD. By increasing the
coupling strength $g,$ we find that the peak shifts to the right,
while the linewidth as well as the amplitude are unchanged. This
suggests that the energy levels of the DQD are shifted so that the
energy splitting is widened under the effect of the NAMR. However,
these changes do
not involve the absorption of any NAMR phonon. As demonstrated in Fig.~\ref%
{backactionNAMR}(b), the frequency shift increases with the square of the
coupling strength between the DQD and the NAMR, consistent with $\chi
=g^{2}/\delta $ in Eq.~(\ref{eq:effective_hamiltonian}).

\subsection{QPC-induced backaction in the current correlation spectrum of
the QPC \label{sec:backaction}}

The backaction on the DQD due to measurement by the QPC is illustrated in
Fig.~\ref{backaction} when the NAMR is practically in its ground state with $%
\langle n\rangle =0.02$. There is no backaction effect when the bias
voltage $V_{\mathrm{QPC}}$ across the QPC is less than the energy
difference between the two DQD eigenstates, \cite{You2010PRB} i.e.,
$eV_{\mathrm{QPC}}<\delta _{0}$. At $eV_{\mathrm{QPC}}=2\pi \times
0.5$~GHz $>\delta _{0}$, for
example, 
a single peak located at $\omega \sim 2\pi \times 0.4$~GHz appears. When $V_{%
\mathrm{QPC}}$ is increased, we find that the linewidth of the
spectrum becomes broadened, which results from $\gamma _{0}=\gamma
_{1}+\gamma _{d}/2$ where $\gamma _{1}$ is proportional to the bias
voltage. Physically, the broadening results from more frequent state
transitions in the DQD induced by the backaction from the QPC when a
larger bias voltage is applied across the QPC.

Dissipations in the NAMR due to the environment have been neglected in our
analysis. Dissipation in an NAMR (see Ref.~%
\onlinecite{Tamayo2005JAP-Cleland2002JAP}) can be expressed as
$\gamma _{m}=\omega _{m}/Q$ with a quality factor $Q$. However, even
for the NAMR-DQD coupling discussed above (e.g., $2\pi \times
30$~MHz), the dissipation of the NAMR is still very small: $\gamma
_{m}/g\sim 10^{-4}$ for an experimentally achievable quality factor
\cite{Huttel2009NanoLett} $Q=10^{5}$. This justifies neglecting the
dissipations of the NAMR due to other environmental effects in our
calculations.

For a QPC, a Kondo-like model was proposed in
Ref.~\onlinecite{Meir2002PRL}, which is similar to the Kondo problem
in a single quantum dot coupled to two leads where the spin degree
of freedom plays an important role. Here, as in
Refs.~\onlinecite{Gurvitz1997PRB} and~\onlinecite{Engel2004PRL}, the
QPC we used is simply modeled as a tunneling junction, and the spin
degree of freedom does not affect its performance. In addition, it
should be noted that the Kondo effect in the DQD can be avoided
here. In fact, the present setup involves no reservoirs (leads)
coupled to the DQD because the coupling between the DQD and the
reservoirs is tuned to zero or is negligibly small. Nevertheless,
the Kondo effect in a DQD needs a strong coupling between the DQD
and the reservoirs in addition to other requirements.

In practice, there are finite cross-capacitive couplings among
various gate electrodes, which affect the whole system. However,
because the coupling between the NAMR and the DQD is in the
dispersive regime, the effect of the NAMR on varying the parameters
of the DQD is small. As for the cross-capacitive couplings in the
DQD system, the experiment in
Ref.~\onlinecite{Hu-Lieber-Marcus2007Nnanotech} showed that the
effect of the cross-capacitive couplings can be canceled by
adjusting the plunger voltage of the detector during sweeps of the
DQD plunger voltages. In our proposed setup in
Fig.~\ref{NAMR-DQD-QPC model}(b), more gate electrodes are
introduced. This will enhance the tunability of the DQD system to
achieve the needed parameters of the system.

\section{Conclusions}

\label{conclusions}

We have proposed an approach to study quantum behaviors of an NAMR
by coupling it indirectly to a QPC as a charge detector via a DQD
serving as a quantum transducer. By detecting the current
correlation spectrum of the charge detector, quantum behaviors of
the NAMR can be observed. It provides interesting insight on the
quantum system as well as dynamics of these backaction effects
induced by an act of measurement, which necessarily perturbs the
system being measured. More importantly, the cooling of the NAMR
down to the quantum regime can be verified. In the quantum regime,
NAMR-phonon-induced shifts (an analog to the Stark shift) of DQD
energy levels as well as their relations with coupling strength
between the NAMR and the DQD are demonstrated. Backaction effects
from the charge detector are also explained.

\section*{Acknowledgements}

This work was supported by the National Basic Research Program of
China Grant No. 2009CB929300, the National Natural Science
Foundation of China Grant No. 91121015, and the Research Grant
Council of Hong Kong SAR under Project No. 5009/08P.

\end{document}